\documentclass[epj,numbook]{svjour}
\begin{document}
\title{Casimir amplitudes in a quantum spherical model
with long-range interaction}
\author{H. Chamati\inst{1} \and D. M. Danchev\inst{2} \and
N. S. Tonchev\inst{1}}
\institute{Georgy Nadjakov Institute of Solid State Physics - BAS,
Tzarigradsko chauss\'{e}e 72, 1784 Sofia, Bulgaria \and Institute of
Mechanics - BAS, Acad. G. Bonchev St. bl. 4, 1113 Sofia, Bulgaria}
\date{\today}
\abstract{
A $d$-dimensional quantum model system confined to a general
hypercubical geometry with linear spatial size $L$ and ``temporal
size'' $1/T $ ( $T$ - temperature of the system) is considered in the
spherical approximation under periodic boundary conditions. For a film
geometry in different space dimensions $\frac 12\sigma <d<\frac
32\sigma $ , where $0<\sigma \leq 2$ is a parameter controlling the
decay of the long-range interaction, the free energy and the Casimir
amplitudes are given. We have proven that, if $d=\sigma$, the Casimir
amplitude of the model, characterizing the leading temperature
corrections to its ground state, is $\Delta =-16\zeta(3)/[5\sigma(4\pi
)^{\sigma/2}\Gamma (\sigma /2)]$. The last implies that the universal
constant $\tilde{c}=4/5$ of the model remains the same for both short,
as well as long-range interactions, if one takes the normalization
factor for the Gaussian model to be such that $\tilde{c}=1$. This is a
generalization to the case of long-range interaction of the well known
result due to Sachdev. That constant differs from the corresponding
one characterizing the leading finite-size corrections at zero
temperature which for $d=\sigma=1$ is $\tilde c=0.606$.
\PACS{ {05.70.Jk}{Critical Phenomena}
\and {64.60.i}{General studies of phase transitions}}}
\maketitle

\section{Introduction}\label{Intr}
The confinement of quantum mechanical vacuum fluctuations of the
electromagnetic field causes long-ranged forces between two
conducting uncharged plates which is known as (quantum mechanical)
Casimir effect \cite{C48,Most97}. The confinement of {\it critical
fluctuations} of an order parameter field induces long-ranged forces
between the surfaces of the film \cite{Fisher78,Krech94}. This is
known as ``statistical-mechanical Casimir effect''. The Casimir
force in statistical-mechanical systems is characterized by the
excess free energy due to the {\it finite-size contributions} to the
free energy of the system. In the case of a film geometry $L\times
\infty ^2$, and under given boundary conditions $\tau $ imposed across
the direction $L$,  the Casimir force is defined as
\begin{equation}
F_{{\rm Casimir}}^\tau (T,L)=-\frac{\partial f_\tau ^{{\rm ex}}(T,L)}{
\partial L},  \label{eq1}
\end{equation}
where $f_\tau ^{{\rm ex}}(T,L)$ is the excess free energy
\begin{equation}
f_\tau ^{{\rm ex}}(T,L)=f_\tau (T,L)-Lf_{{\rm bulk}}(T).  \label{eq2}
\end{equation}
Here $f_\tau (T,L)$ is the full free energy per unit area and per
$k_BT$, and $f_{{\rm bulk}}(T)$ is the corresponding bulk free
energy density.

The full free energy of a $d$-dimensional critical system in the form
of a film with thickness $L$, area $A$, and boundary conditions $a$
and $b$ on the two surfaces, at the bulk critical point $T_c$, has the
asymptotic form
\begin{eqnarray}
f_{a,b}(T_c,L|d)&\cong& Lf_{{\rm bulk}}(T_c|d)+f_{{\rm
surface}}^a(T_c|d) \label{eq3}\\
& &+f_{{\rm
surface}}^b(T_c|d)+L^{-(d-1)}\Delta _{a,b} (d) + \cdots \nonumber
\end{eqnarray}
as $A\to \infty $, $L\gg 1$. Here $f_{{\rm surface}}$ is the surface
free energy contribution and $\Delta _{a,b} (d)$ is {\it the amplitude
of the Casimir interaction}. The $L$ dependence of the Casimir term
(the last one in Eq.~(\ref{eq3})) follows from the scale invariance of
the free energy and has been derived by Fisher and de
Gennes~\cite{Fisher78}. The amplitude $\Delta _{a,b} (d)$ is {\it
universal}, depending on the bulk universality class and the
universality classes of the boundary
conditions~\cite{Krech94,Krech92}.

Equation~(\ref{eq3}) is valid for both {\it fluid and magnetic systems
at criticality}. Prominent examples are, e.g., one-component fluid at
the liquid-vapour critical point, the binary fluid at the consolute
point, and liquid $^4$He at the $\lambda $ transition point
~\cite{Krech94}. The boundaries influence the system to a depth given
by the bulk correlation length $\xi _\infty (T)\sim |T-T_c|^{-\nu }$,
where $\nu $ is its critical exponent. When $\xi
_\infty (T)\ll L$ the Casimir force, as a {\it fluctuation induced
force} between the plates, is negligible. The force becomes
long-ranged when $\xi _\infty (T)$ diverges near {\it and} below the
bulk critical point $T_c$ in an ${\cal O}(n)$, $n\geq 2$ model system
in the absence of an external magnetic field~\cite
{Danchev96,Danchev98}. Therefore in statistical-mechanical systems one
can turn on and off the Casimir effect merely by changing, e.g., the
temperature of the system.

The temperature dependence of the Casimir force for two-dimensional
systems has been investigated exactly only on the example of Ising
strips~\cite{Evans94}. In ${\cal O}(n)$ models for $T>T_c$ the
temperature dependence of the force has been considered
in~\cite{Krech92}. The only example where it is investigated
exactly as a function of both the temperature and magnetic field
scaling variables is that of the three-dimensional spherical model
under periodic boundary conditions~\cite{Danchev96}. There exact results for
the Casimir force between two walls with a finite separation in a
$L\times\infty^2$ mean-spherical model have been derived. The force is
consistent with an {\it attraction} of the plates confining the
system.

The most of the results available at the moment are for the Casimir
amplitudes. They are obtained for $d=2 $ by using conformal-invariance
methods for a large class of models~\cite{Krech94}. For $d\neq 2$
results for the amplitudes are available via field-theoretical
renormalization group theory in $4-\varepsilon$
dimensions~\cite{Krech94,Krech92,Eisen95}, Migdal-Kadanoff real-space
renormalization group methods~\cite{Indeku}, and, relatively recently,
by Monte Carlo methods~\cite{krech96}. In addition to the flat
geometries recently some results about the Casimir amplitudes between
spherical particles in a critical fluid have been derived
too~\cite{Eisen95,HSED98}. For the purposes of experimental
verification that type of geometry seems more perspective.

It should be noted that in contrast with the quantum mechanical
Casimir effect, that has been tested experimentally with high accuracy
\cite{L97}, the statistical-mechanical Casimir effect lacks so far a
satisfactory experimental verification (for comments on the specific
difficulties that the experiment stacks with see, e.g. \cite{HSED98}).

In recent years there has been a renewed
interest~\cite{Sachdev96,Sondhi97} in the theory of zero-temperature
quantum phase transitions. In contrast to temperature driven critical
phenomena, these phase transitions occur at zero temperature as a
function of some non-thermal control parameter, say $g$, (or a
competition between different parameters describing the basic
interaction of the system), and the relevant fluctuations are of
quantum rather than thermal nature. In the present article we consider
a statistical-mechanical Casimir effect when critical quantum
fluctuations play an essential role.

It is well known from the theory of critical phenomena that for
temperature driven phase transitions quantum effects are unimportant
near critical points with $T_c>0$. It could be expected, however, that
at rather low (as compared to characteristic excitations in the
system) temperatures, the leading $T$ dependence of all observables is
specified by the properties of the zero-temperature critical point,
say at $g_c$. The {\it dimensional crossover rule } asserts that the
critical singularities with respect to $g$ of a $d$-dimensional
quantum system at $T=0$ and around $g_c$ are {\it formally} equivalent
to those of a classical system with dimensionality $d+z$ ($z$ is the
dynamical critical exponent) and critical temperature $ T_c>0$. This
makes it possible to investigate low-temperature effects (considering
an effective system with $d$ infinite spatial and $z$ finite temporal
dimensions) in the framework of the theory of finite-size scaling
(FSS). This theory has been applied to explore the low-temperature
regime in quantum systems~\cite{Sachdev96,Sondhi97,Chakra89}, when the
properties of the thermodynamic observables in the finite-temperature
quantum critical region have been the main focus of interest.

In this paper a theory of the scaling properties of the free energy
and Casimir amplitudes of a quantum spherical model~\cite{Vojta96}
with nearest-neighbor and some special cases of long-range
interactions (decreasing at long distances $r$ as $1/r^{d+\sigma }$)
is presented. These interactions enter the exact expressions for the
free energy only through their Fourier transform wich leading
asymptotic is $U(q)\sim q^{\sigma^*}$, where $\sigma^*=min(\sigma,2)$
\cite{joyce72}. As it was shown for bulk systems by renormalization
group arguments $\sigma\ge 2$ corresponds to the case of finite
(short) range interactions, i.e. the universality class then does not
depend on $\sigma$ \cite{FMN72}. Values satisfying $0<\sigma<2$
correspond to long-range interactions and the critical behaviour
depends on $\sigma$. On the above reasoning one usually considers the
case $\sigma>2$ as uninteresting for critical effects, even for the
finite-size treatments \cite{fisher86}. However recent Monte Carlo
results suggest that it might well not be the case at least for
continuous Ising model \cite{BD99}. There Bayong and Diep state that
for $d=2$ the critical exponents does not depend on $\sigma$ and reach
their short-range values for $\sigma\ge 3$. On the basis of that
result it seems that for finite-size systems the case $\sigma>2$ is
nontrivial. Since, up to the authors knowledge that is the only
example where $\sigma>2$ is of interest for studying critical
properties, here we will consider only the case $0<\sigma\le 2$.

The investigation of the Casimir effect in a classical system with
long-range interaction possesses some peculiarities in comparison with
the short-range system. Due to the long-range character of the
interaction there exists a natural attraction between the surfaces
bounding the system. One easily can estimate that in the ordered state
the $L$-dependent part of the excess free energy that is raised by the
direct inter-particle (spin) interaction is of order of
$L^{-\sigma+1}$. In the critical region one still has some effect
 stemming from that interaction on the background of which develops
the fluctuating induced new attraction between the surfaces which is
in fact the critical Casimir force. In the definition (\ref{eq1}) used
here, that is the common one when one considers short-range systems,
these both effects are superposed simultaneously. Therefore, here,
generally speaking, one should expect a crossover from a regime
governed by the critical Casimir force (in the sense of a fluctuation
induced force; it is of the order of $L^{-d}$, see Eq. (\ref{eq3})) to
the one govern by the direct attraction (of the order of
$L^{-\sigma}$; note that if $d=\sigma$ they will of the same order
being dominating in different temperature regions). An interesting
case when forces of similar origin are acting simultaneously is that
one of the wetting when the wetting layer is nearly critical and
intrudes between two noncritical phases if one takes into account the
effect of long-range correlations and that one of the long-range van
der Waals forces \cite{NI85,note1}.

In quantum systems additional new features will be observed since the
``temporal direction'' corresponds formally to a short-range type
interaction in the corresponding classical analog of the system, i.e.
one unavoidable has ``anisotropy'' in the spectrum of a quantum system
with long-range interactions. The effects depend on which dimension --
the temporal or the spatial one -- is the finite one. When the spatial
dimension is finite we only mention here and will demonstrate in the
current article, that an effect similar to that for the classical
systems exists. If the finite dimension is the temporal one such
effect will not be observed since then the long-range interaction is
inter-surficial and is not effected directly by the ``finite size'' of
the system.

The plan of the paper is as follows. In Section 2 we define some
generalization of Casimir amplitudes in quantum systems and present
some hypotheses for the corresponding excess free energies. Then in
Section 3 we give a brief review of the model and the basic equations
for the free energy and the quantum spherical field in the case of
periodic boundary conditions. Since we make use of the ideas of the
FSS theory, the bulk system in the low-temperature region is
considered as an effective ($d+z$) dimensional classical system with
$z$ finite (temporal) dimensions. This is done to make possible a
comparison with other results based on the spherical type
approximation, e.g., in the framework of the spherical model and the
QNL$\sigma $M in the limit $n\to \infty $. The scaling forms for the
excess free energy, the spherical field equation and the Casimir force
are derived for a $\frac 12\sigma <d<\frac 32\sigma $ dimensional
system with a film geometry in Section~\ref{Free}. In
Section~\ref{Casimir} we present some results for the Casimir
amplitudes in the case of short-range interactions and in some special
cases of long-range interactions. The paper closes with concluding
remarks given in Section~\ref{Concl}.

\section{Casimir amplitudes in critical quantum systems}

Let us consider a quantum system with a film geometry $L\times
\infty ^{d-1}\times L_\tau $, where $L_\tau \sim\hbar /(k_BT)$ is
the ``finite-size'' in the temporal (imaginary time) direction and
let us suppose that {\it periodic
boundary conditions}
are imposed across the finite space
dimensionality $L$ (in the remainder
we will set $\hbar =k_B=1$).
Let $f(T,g,H;L|d)$ be the free energy density of
this system. Then, according to the dimensional crossover rule, the
Privman - Fisher hypothesis \cite{fisher84} for finite classical
systems and Eq. (\ref{eq2}) in the quantum case one could state that
\begin{equation}
\frac1L f^{{\rm ex}}(T,g ,H;L|d)=\left( TL_\tau \right)
L^{-(d+z)}X_{{\rm ex}}^{{\rm u}}(x_1,x_2,\rho |d), \label{hypot1}
\end{equation}
with scaling variables
\begin{equation}
x_1=L^{1/\nu } \delta g,\ x_2=hL^{\Delta
/\nu } \ \mbox{and}\ \rho = L^z/L_\tau .
\end{equation}
Here $\Delta$ and $\nu$ are the
usual critical exponents of the bulk model, $h$ is a properly normalized
external magnetic field $H$, $\delta g \sim g-g_c$, and $ X_{{\rm
ex}}^{{\rm u}}$ is the universal scaling function of the excess free
energy. According to the definition (\ref{eq1}) of the Casimir force,
one obtains immediately
\begin{equation}
F_{{\rm Casimir}}^{d}(T,g ,H;L)=\left( TL_\tau \right)
L^{-(d+z)}X_{{\rm Casimir}}^{{\rm u}}(x_1,x_2,\rho |d),
\label{defCasimir}
\end{equation}
where the {\it universal} scaling functions of the Casimir force
$X_{{\rm Casimir}}^{{\rm u}}(x_1,x_2,\rho |d)$ is related to the one
of the excess free energy $X_{{\rm ex}}^{{\rm u}}\equiv X_{{\rm
ex}}^{{\rm u} }(x_1,x_2,\rho |d)$ by
\begin{eqnarray}
X_{{\rm Casimir}}^{{\rm u}}(x_1,x_2,\rho |d)&=&-(d+z)X_{{\rm
ex}}^{{\rm u} }-\frac 1\nu x_1\frac{\partial X_{{\rm ex}}^{{\rm
u}}}{\partial x_1}\nonumber\\ && +\frac\Delta \nu
x_2\frac{\partial X_{{\rm ex}}^{{\rm u}}}{\partial x_2}+z\rho
\frac{
\partial X_{{\rm ex}}^{{\rm u}}}{\partial \rho }.
\label{DefXCasimir}
\end{eqnarray}
It follows from Eq. (\ref{defCasimir}) that depending on the scaling
variable $\rho $ one can consider the general case of Casimir
amplitudes
\begin{equation}
\Delta _{{\rm Casimir}}^{{\rm u}}\left( \rho |d \right) = X_{{\rm
Casimir}}^{{\rm u}}\left( 0,0,\rho |d \right) . \label{DeltaDef}
\end{equation}
The classical amplitudes $\Delta_{a,b} (d)$ (for $(a,b)\equiv$
periodic boundary conditions) introduced by Eq. (\ref{eq3}) are
particular cases of $\Delta _{{\rm Casimir}}^{{\rm u}}\left( \rho |d
\right)$ for $\rho =0$, i.e. $T=0$ (we remind that our system,
according to the dimensional crossover rule, is formaly equivalent to
a $d+z$ dimensional classical one).

In addition to the above ``usual'' excess free energy and Casimir
amplitudes, denoted by the superscript ``$u$'', one can define, in a
full analogy with what it has been done above, ``{\it temporal excess
free energy density''} $f_{{\rm t}}^{{\rm ex}}$,
\begin{equation}
f_{{\rm t}}^{{\rm
ex}}(T,g,H|d)=f(T,g,H;\infty|d)-f(0,g,H;\infty|d) \label{deffext}
\end{equation}
and ``{\it temporal Casimir amplitudes}''
\begin{equation}
f_{{\rm t}}^{{\rm ex}}(T,g_c,0|d) =TL_\tau ^{- d/z} \Delta _{{\rm
Casimir}}^{{\rm t}}\left( d\right). \label{Deltatdef}
\end{equation}
Whereas the ``usual'' amplitudes characterize the leading $L$
corrections to the bulk free energy density at the critical point, the
``temporal amplitudes'' determine the leading temperature-dependent
corrections to the ground state energy of an {\it infinite} system at
its quantum critical point $g_c$.

If in Eq. (\ref{Deltatdef}) the quantum parameter $g$ is in the
vicinity of $g_c$, then one expects
\begin{equation}
f_{{\rm t}}^{{\rm ex}}(T,g,H) =TL_\tau ^{- d/z}X_{{\rm ex}}^{{\rm
t}}\left( x_1^t,x_2^t|d\right), \label{defxextgen}
\end{equation}
i.e. instead of the amplitude $\Delta _{{\rm Casimir}}^{{\rm
t}}\left( d \right) $ one has a scaling function $X_{{\rm
ex}}^{{\rm t}}\left( x_1^t,x_2^t|d \right) $ which is the
corresponding analog of $X_{{\rm ex}}^{{\rm u}}(x_1,x_2,\rho |d)$.
The scaling variables now are
\begin{equation}
x_1^t=L^{1/\nu z } \delta g \ \mbox{and} \
x_2^t=hL^{\Delta /\nu z }.
\end{equation}
Obviously
\begin{equation}
\Delta _{{\rm Casimir}}^{{\rm t}}\left( d\right)
=
X_{{\rm ex}}^{{\rm t}}\left(0,0|d\right). \label{defCastgen}
\end{equation}
Let us finally note that if $z=1$ the temporal excess free energy
introduced here coincides, up to a (negative) normalization factor,
with the proposed by Neto and Fradkin \cite{CN93} definition of the
non-zero temperature generalization of the $C$-function of
Zamolodchikov.

Now we pass to study the quantities introduced above on the example of
one exactly solvable model.

\section{The model}
\label{Model}

The model we consider is described by the Hamiltonian~\cite{Vojta96}
\begin{equation}
{\cal H}=\frac 12g\sum_\ell {\cal P}_\ell ^2-\frac 12\sum_{\ell \ell
^{\prime }}{J}_{\ell \ell ^{\prime }}{\cal S}_\ell {\cal S}_{\ell^{
\prime}}+\frac 12\mu \sum_\ell{\cal S}_\ell^2-H\sum_\ell{\cal S}_\ell ,
\label{eq4}
\end{equation}
where ${\cal S}_\ell $ are spin operators at site $\ell $. The
operators ${\cal P}_\ell $ play the role of ``conjugated'' momenta
(i.e. $[{\cal S}_\ell ,{\cal S}_{\ell ^{\prime }}]=0$, $[{\cal P}_\ell
,{\cal P}_{\ell^{\prime }}]=0$, and $[{\cal P}_\ell ,{\cal S}_{\ell
^{\prime }}]=i\delta_{\ell \ell ^{\prime }}$, with $\hbar =1$). The
coupling constant $g$ measures the strength of the quantum
fluctuations (below it will be called quantum parameter), $H$ is an
ordering magnetic field, and the spherical field $\mu $ is introduced
so as to ensure the constraint
\begin{equation}
\sum_\ell \left\langle {\cal S}_\ell ^2\right\rangle =N.  \label{eq5}
\end{equation}
Here $N$ is the total number of quantum spins located at sites ``$\ell
$'' of a finite hypercubical lattice $\Lambda $ of size $L_1\times
L_2\times\cdots \times L_d=N$ and $\left\langle \cdots
\right\rangle $ denotes the standard thermodynamic average taken
with the Hamiltonian ${\cal H}$. In Ref. \cite{Vojta96}, the
equivalence of the model~(\ref{eq4}) and the quantum ${\cal O}(n)$
nonlinear sigma model in its large $n$-limit is shown.

Let us note that in the last few years an increasing interest in the
{\it spherical approximation} (or large $n$-limit), generating
tractable models of quantum critical phenomena, has been observed
~\cite {Vojta96,Tu94,Nieu95,Nieu97,Chamati97,chamati97}. There are
different possible ways of quantization of the spherical constraint.
In general they lead to {\it different universality classes} at the
quantum critical point~\cite{Vojta96,Tu94,Nieu95,Nieu97}. The
commutation relations for the operators ${\cal S}_\ell$ and ${\cal
P}_\ell$ together with the kinetic term in the Hamiltonian~(\ref{eq4})
do not describe quantum Heisenberg-Dirac spins but quantum rotors as
is pointed out in Ref.~\cite{Vojta96}. Since the quantum rotors model has
been widely exploited in the field of high-temperature superconductivity
(see, e.g. \cite{Sachdev96} and references therein) we hope that the
treatment of the model (\ref{eq4}) presented below might be of some
interest to those problems.

For nearest neighbour interaction  different low temperature
regimes and finite-size scaling properties of the model are investigated
in Ref. \cite{Chamati97}.

The free energy of the model in a finite region $\Lambda $ under
periodic boundary conditions applied across the finite dimensions has
the form~\cite {chamati97}
\begin{eqnarray}
\beta f_\Lambda \left( \beta ,g,H\right)&=&\sup_\mu \left\{ \frac 1N\sum_q\ln
\left[ 2\sinh \left(\frac 12\beta \omega \left( {q};\mu \right) \right)
\right]\right.\nonumber\\
& & \left.-\frac \mu 2\beta -\frac{\beta g}{2\omega ^2\left( 0;\mu
\right) } H^2\right\} . \label{eq6}
\end{eqnarray}
Here the vector $q$ has the components $\left\{ \frac{2\pi
n_1}{L_1},\cdots ,\frac{2\pi n_d}{L_d}\right\}$, $n_j\in \left\{
-\frac{L_j-1}2,\cdots ,\frac{L_j-1}2\right\} $ for $L_j$ odd
integers, and $\beta $ is the inverse temperature (with $k_B=1$).
In~(\ref{eq6}) the spectrum is $\omega
^2\left( {q};\mu \right) =g\left( \mu +U({q})\right) $ with
$U({q})\cong {J}|{q} |^\sigma $, $0<\sigma \le 2$.
 In the above expressions $U({q})$ is the Fourier transform
of the interaction matrix where the energy scale has been fixed so
that $U(0)=0$. The supremum in Eq.~(\ref{eq6}) is attained at the
solutions of the mean-spherical constraint, Eq.~(\ref{eq5}), that
reads
\begin{equation}
1=\frac tN\sum_{m=-\infty }^\infty \sum_q\frac 1{\phi
+U({q})/{J}+b^2m^2}+\frac{h^2}{\phi ^2}, \label{eq7}
\end{equation}
where we have introduced the notations: $b=(2\pi t)/\lambda $,
$\lambda =\sqrt{g/J}$ is the normalized quantum parameter, $t=T/{J}$ -
the normalized temperature, $h=H/J$ - the normalized magnetic field,
and $\phi =\mu /{J}$ is the scaled spherical field. Eqs.~(\ref{eq6})
and~(\ref{eq7}) provide the basis for studying the critical behaviour
of the model under consideration.

In the thermodynamic limit it has been shown~\cite{Vojta96} that for
$d>\sigma$ the long-range order exists at finite temperatures up to a
given critical temperature $t_c(\lambda)$. Here we shall consider the
{\it low-temperature region} for
$\frac{1}{2}\sigma<d<\frac{3}{2}\sigma$. We remind that
$\frac{1}{2}\sigma$ and $\frac{3}{2}\sigma$ are the lower and the
upper critical dimensions, respectively, for the quantum critical
point of the considered system.

\section{Scaling form of the excess free energy and the Casimir force
at low temperatures}\label{Free}
For a system with a film geometry
$L\times \infty ^{d-1}\times L_\tau $ (where $\frac 12\sigma <d<\frac
32\sigma $), after taking the limits $L_2\to \infty ,\cdots
,L_d\to\infty$ in Eq.~(\ref{eq6}) with $ L_1=L $, we receive the
following expression for the full free energy density (see
Appendix~\ref{appendixA})
\begin{eqnarray}
&& f(t,\lambda ,h;L|d,\sigma )/J= -\frac{h^2}{2\phi }-\frac \phi 2
\nonumber\\
&& \ \ \ \ \ \ +\frac{\lambda k_d}{2d}x_D^d\left( x_D^\sigma
+\phi\right)^{\frac12}\ _2F_1\left( 1,-\frac 12,1+\frac d\sigma
,\frac{x_D^\sigma }{ x_D^\sigma +\phi }\right)\nonumber \\ &&
\ \ \ \ \ \ -\frac \lambda 4\frac{\sigma
L^{-(d+\frac\sigma2)}}{\left(4\pi\right)^{\frac d2} }\sum_{n=1}^\infty
\int_0^\infty dx x^{-\frac\sigma4-\frac d2-1}\exp \left( -\frac{n^2}{
4x}\right)\nonumber \\ &&\ \ \ \ \ \ \ \ \ \ \ \ \ \ \ \ \ \ \ \ \ \ \
\ \times G_{\frac\sigma 2,1-\frac
\sigma 4}\left(-x^{\sigma /2}L^\sigma\phi \right) \nonumber\\
&& \ \ \ \ \ \ \ -\frac\lambda\sigma\frac{k_d}{\sqrt{\pi
}}\Gamma\left(\frac d\sigma
\right)\phi^{\frac d\sigma+\frac12}\sum_{m=1}^\infty\frac{K_{\frac d\sigma
+\frac 12}\left( m\frac \lambda t\phi ^{\frac 12}\right)}{ \left( m\frac
\lambda {2t}\phi ^{\frac 12}\right) ^{\frac d\sigma +\frac 12}}\nonumber \\
&& \ \ \ \ \ \ -\lambda
\sqrt{2}\frac{L^{-(d+\frac\sigma2)}}{(2\pi )^{\frac{d+1}2}}
\!\sum_{n=1}^\infty\!\sum_{m=1}^\infty\int_0^\infty \frac{dz}{
m^dz^{\frac32}}\!{\cal F}_{\frac d2-1,\sigma }\left(\frac{z}{m^\sigma
}\right)\nonumber\\ && \ \ \ \ \ \ \ \ \ \ \ \ \ \ \ \ \ \ \ \ \ \ \ \
\times\exp\left[-zL^\sigma\phi-\frac{n^2}{4zL^\sigma
}\left(\frac\lambda t\right)^2\right] \!.
 \label{eq8}
\end{eqnarray}
Here $k_d^{-1}=\frac 12(4\pi )^{\frac d2}\Gamma (d/2)$, $x_D$ is the
radius of the sphericalized Brillouin zone,
\begin{equation}
G_{\alpha ,\beta }\left( t\right) =\frac 1{\sqrt{\pi }}\sum_{k=0}^\infty
\frac{\Gamma \left( k+1/2\right) }{\Gamma \left( \alpha k+\beta \right) }
\frac{t^k}{k!}  \label{G}
\end{equation}
(it was introduced in Ref. \cite{Chamati}),
\begin{equation}
{\cal F}_{\nu ,\sigma }\left( y\right) =\int_0^\infty x^{\nu +1}J_\nu \left(
x\right) \exp \left( -yx^\sigma \right) dx  \label{F}
\end{equation}
and $K_\nu (x),$ and $J_\nu (x)$ are the MacDonald and Bessel
functions, respectively. The main advantage of the above expression,
despite of its complicated form in comparison with Eq. (\ref{eq6}), is
the simplified dependence on the size $L$ which now enters only via
the arguments of some functions. This gives us the possibility, as it
is explained below, to obtain the scaling functions of the excess free
energy and the Casimir force.

In Eq.~(\ref{eq8}) $\phi $ is the solution of the corresponding
spherical field equation that follows by requiring the partial
derivative of the r.h.s. of Eq.~(\ref{eq8}) with respect to $\phi $ to
be zero. The bulk free energy $f_{{\rm bulk}}(t,\lambda ,h|d,\sigma )$
results from $f(t,\lambda ,h;L|d,\sigma )$ by merely taking the limit
$L\to\infty $ in it. Let us denote the solution of the corresponding
bulk spherical field equation by $\phi _\infty $. Then for the excess
free energy it is possible to obtain from Eqs. (\ref{eq2}) and (\ref{eq8}),
in full accordance with Eq.
(\ref{hypot1}), the finite size scaling form
\begin{equation}
\frac1L f^{{\rm ex}}(t,\lambda ,h;L|d,\sigma )=\left( TL_\tau
\right) L^{-(d+z)}X_{{\rm ex}}^{{\rm u}}(x_1^u,x_2^u,\rho |d,\sigma ),
\label{eq10}
\end{equation}
with scaling variables
\begin{equation}
x_1^u=L^{1/\nu }\left(\lambda^{-1}
-\lambda _c^{-1}\right), \  x_2^u=hL^{\Delta /\nu } \ \mbox{and} \ \rho
=L^z/L_\tau,
\end{equation}
with $L_\tau =\lambda /t$.
Here the critical value
of $\lambda =\lambda _c$ is
\begin{equation}
\lambda _c^{-1}=\frac 12(2\pi )^{-d}\int d^d{q}(U({q})/{J})^{-\frac 12},
\label{eq9}
\end{equation}
and $\nu ^{-1}=d-\frac 12\sigma $, $\Delta /\nu =\frac 12\left(
d+\frac 32\sigma \right) ,$ and $z=\frac 12\sigma $ are the critical
exponents of the model~\cite{Vojta96}. In Eq.~(\ref{eq10}) the {\it
universal} scaling function $X_{{\rm ex}}^{{\rm u}}(x_1^u,x_2^u,\rho
|d,\sigma )$ of the excess free energy has the form
\begin{eqnarray}
&&X_{{\rm ex}}^{{\rm u}}(x_1^u,x_2^u,\rho |d,\sigma ) =\frac 12x_1^u\left(
y_\infty-y_0 \right) +\frac 12(x_2^u)^2\left( \frac 1{y_\infty}-\frac
1{y_0 }\right)\nonumber \\ &&
 \ \ \ \ \ \ \ \ \ \ \ -\frac{k_d}{4\sqrt{\pi }\sigma }\Gamma \left(
\frac d\sigma \right) \Gamma
\left( -\frac d\sigma -\frac 12\right) \left( y_0^{\frac d\sigma +\frac
12}-y_\infty ^{\frac d\sigma +\frac 12}\right) \nonumber \\ &&\ \ \ \
 \ \ \ \ \ \ \ -\frac{k_d}{\sigma \sqrt{\pi }}\Gamma \left( \frac
d\sigma \right)
\sum_{m=1}^\infty \left[ \frac{\left( 2y_0\right) ^{\frac d\sigma +\frac
12}K_{\frac d\sigma +\frac 12}\left( m\frac{\sqrt{y_0}}\rho \right) }
{\left( m\frac{\sqrt{y_0}}\rho\right) ^{\left( \frac d\sigma +\frac
12\right) }}\right. \nonumber \\ &&
 \ \ \ \ \ \ \ \ \ \ \ \ \ \ \ \ \ \ \ \ \ \ \ \ \ \ \ \ \
-\left. \frac{\left(2y_\infty\right) ^{\frac d\sigma +\frac 12}
K_{\frac d\sigma +\frac12}\left( m\frac{\sqrt{y_\infty}}\rho\right)
}{\left( m\frac{\sqrt{y_\infty}}\rho\right)
^{\left( \frac d\sigma +\frac 12\right) }}\right] \nonumber \\
&& \ \ \ \ \ \ \ \ \ \ \ -\frac 14\frac \sigma {\left( 4\pi \right)
^{\frac d2}}\sum_{n=1}^\infty\int_0^\infty x^{-\frac\sigma4-\frac d2-1}
\exp \left( -\frac{n^2}{4x}\right)\nonumber\\
&& \ \ \ \ \ \ \ \ \ \ \ \ \ \ \ \ \ \ \ \ \ \ \ \ \ \ \ \ \ \times
G_{\frac\sigma 2,1-\frac \sigma 4}
\left( -x^{\frac\sigma2}y_0\right) dx\nonumber \\
&& \ \ \ \ \ \ \ \ \ \ \ -\frac{\sqrt{2}}{(2\pi
)^{\frac{d+1}2}}\!\sum_{n=1}^\infty
\!\sum_{m=1}^\infty \int_0^\infty \frac{dz}{m^dz^{\frac32}}\!{\cal F}
_{\frac d2-1,\sigma }\left(\frac z{m^\sigma }\right)\nonumber\\
&& \ \ \ \ \ \ \ \ \ \ \ \ \ \ \ \ \ \ \ \ \ \ \ \ \ \ \ \ \
\times\exp
\left[ -zy_0-\frac{n^2}{4z\rho ^2}\right]
 \label{Xscaling}
\end{eqnarray}
(see Appendix~\ref{appendixA} for details of the calculations), where
$y_0=\phi L^\sigma $ and $y_\infty =\phi _\infty L^\sigma .$

By direct
evaluation of the above expression it is easy to see that below $\lambda_c$
if the finite system is in ordered phase (then $y_0=y_\infty=0$)
$X_{{\rm ex}}^{{\rm u}}\sim L^{-\sigma}$, that reflects the dominating
contribution of the direct inter-spin long-range interaction in that region
\cite{note2}.
As we see from Eq. (\ref{eq10}) one observes a crossover from $L^{-(d+z)}$
behavior, where the fluctuation induced interactions dominate,
to $L^{-\sigma}$ one
where the direct inter-spin interactions become essential.

For the Casimir force one obtains
\begin{equation}
F_{{\rm Casimir}}^{d,\sigma }(T,\lambda ,h;L)=\left( TL_\tau \right)
L^{-(d+z)}X_{{\rm Casimir}}^{{\rm u}}(x_1^u,x_2^u,\rho |d,\sigma ),
\label{eq12}
\end{equation}
where the {\it universal} scaling function of the Casimir force
$X_{{\rm Casimir}}^{{\rm u}}(x_1^u,x_2^u,\rho |d,\sigma )$ is related
to that one of the excess free energy
$X_{{\rm ex}}^{{\rm u}}
\equiv X_{{\rm ex}}^{{\rm u} }(x_1^u,x_2^u,\rho |d,\sigma )$ by Eq.
(\ref{DefXCasimir}).

The above expressions for the scaling functions of the excess free
energy (and the Casimir force) are the most general ones, which
gives the possibility of a general analysis including issues as:
{\it i)} the sign of the Casimir force; {\it ii)} monotonicity of
the Casimir force as a function of the temperature; {\it iii)} the
relation of the excess free energy scaling function to the
corresponding finite-temperature $C$-function and its monotonicity
properties; {\it iv)} finite-system generalization of the
finite-temperature $C$-function, etc.

In the present article we will concentrate on evaluation of the
Casimir amplitudes for some special cases where one can obtain
simple analytical expressions for them.

It is clear that just due to the dimensional crossover rule $L$
plays the same role for the finite system at $t=0$  as $L_\tau$
for the corresponding infinite quantum system. Therefore, by a
symmetry that obviously arises when $\sigma =2,$ one should expect
that the behavior of the two types of amplitudes (``normal'' and
``temporal'') will be essentially the same. We will see that
explicitly below.

For the model we study here one can show that
\begin{eqnarray}
&&X_{{\rm ex}}^{{\rm t}}(x_1^t,x_2^t|d,\sigma )=\frac 12 x_1^t \left(
y_\infty-y_0 \right)+\frac 12 (x_2^t)^2\left( \frac 1{y_\infty}-\frac
1{y_0 }\right)\nonumber \\ &&
 \ -\frac{k_d}{4\sqrt{\pi }\sigma }\Gamma \left(
\frac d\sigma \right) \Gamma
\left( -\frac d\sigma -\frac 12\right) \left(y_0^{(d/z+1)/2}-
y_\infty ^{(d/z+1)/2}\right) \nonumber \\ &&
 \ -\frac{k_d}{\sigma \sqrt{\pi }}\Gamma \left( \frac d\sigma
\right)\left( 2y_0\right) ^{\frac d\sigma+\frac12} \sum_{m=1}^\infty
\frac{K_{\frac d\sigma +\frac 12}\left( m \sqrt{y_0}\right) } {\left( m
\sqrt{y_0}\right) ^{\frac d\sigma +\frac 12}}.
\label{xext}
\end{eqnarray}
Here the scaling variables are defined by
\begin{equation}
x_1^t=L_\tau ^{1/\nu
z}\left(\frac1\lambda-\frac1\lambda_c\right), \ x_2^t=hL_\tau
^{\Delta /z\nu },
\end{equation}
$y_0=L_\tau ^2 \phi_0$ and $y_\infty=L_\tau ^2 \phi_\infty$.
Note that  $y_0$ is
the solution of the corresponding spherical filed equation for the
nonzero-temperature system, whereas $y_\infty$ is the solution for
the zero-temperature ("infinite" in the ``temporal'' dimension)  one.
The direct
evaluation of (\ref{xext}), supposing the finite system in a ordered state,
shows that $ X_{{\rm ex}}^{{\rm t}}$ remains of the same order in that region.
This is exactly the behavior to be expected, as we mentioned in the
Introduction, despite the long-range nature of the interactions, since the
finite ``dimension'' of the system is now the temporal one.

Now we are ready to investigate in a bit more detail the behavior
of the Casimir amplitudes as a function of $d,\sigma $ and $\rho$.

\section{Evaluation of Casimir amplitudes}
\label{Casimir}

In this section we determine the Casimir amplitudes of the model in
the case of short-range interactions at $d=2$ and in the special case
$d=\sigma$ of long-range interactions.

\subsection{``Usual'' Casimir amplitudes}

\subsubsection{Two-dimensional system with short-range interactions
($d=\sigma=2$)}

In this case essential simplifications in the expression for the
Casimir forces can be made. The functions $G_{\alpha ,\beta }$ and
${\cal F}_{\alpha ,\beta }$ used in the general expression of the
free energy~(\ref{eq8}) turn into the explicit forms
\begin{equation}
G_{1,1/2}(z)=\frac 1{\sqrt{\pi }}\exp \left( z\right)  \label{Gsr}
\end{equation}
and
\begin{equation}
{\cal F}_{\alpha ,2}(z)=\left( 2z\right) ^{-\alpha -1}\exp \left( -\frac
1{4z}\right) .  \label{Fsr}
\end{equation}
At the quantum critical point $\lambda =\lambda _c$, $h=0$ this leads
to ( $y_\infty =0$ )
\begin{eqnarray}
X_{ex}^u(0,0,\rho|2,2)&=&-\left(\frac{y_0}{ 2\pi} \right)
^{\frac32}{\sum_{m,n}}^\prime\frac{K_{\frac32}
\left(\sqrt{y_0\left(\frac{n^2}{\rho^2}+m^2\right)
}\right) }{\left( \sqrt{y_0\left(\frac{n^2}{\rho^2}+m^2\right) }\right)
^{\frac32}}
\nonumber \\ &&-\frac 1{12\pi }y_0^{3/2}\label{Xcr},
\end{eqnarray}
where the primed summation over the integers $m$ and $n$ indicates
that the term corresponding to $m=n=0$ is excluded.

Since it is not clear how to obtain an explicit analytical
solution for $y_0$ in a film geometry at nonzero temperature, the
above expression cannot be simplified further, but has to be
analyzed numerically (see, e.g., \cite {Chamati97} for a numerical
analysis of the spherical filed equation). Nevertheless, the above
expression can be significantly simplified at zero temperature.
Then one shows that the solution $y_0$ of the spherical field
equation for the finite system with a film geometry $L\times
\infty \times L_\tau ,$ at zero temperature (i.e. $1\ll L\ll\infty
$, $L_\tau=\infty $) and at the quantum critical point $\lambda
=\lambda _c$, $h=0$ \cite {Chamati97,chamati97} is $y_0=4\ln^2 \left(
\sqrt{5}/2+1/2\right) $. Setting this value of $y_0$ in
~(\ref{Xcr}), taking into account that $K_{3/2}(x)=\sqrt{\pi
/(2x)}\exp (-x)(1+1/x)$, and using the properties of the
polylogarithm functions ${\rm Li}_p(x)$~\cite{Sachdev93}, we
obtain after some algebra that the Casimir amplitude is
\begin{equation}
\Delta _{{\rm Casimir}}^{{\rm u}}\left( 0|2,2\right) =-\frac{2\zeta (3)}{
5\pi }\approx -0.1530. \label{eq14}
\end{equation}
Here $\zeta (x)$ is the Riemann zeta function.

\subsubsection{One-dimensional system with long-range interactions
($d=\sigma=1$)}

We can obtain a relatively simple analytical expression from
Eq.~(\ref{Xscaling}) only in the particular case $\sigma=1$. In
this case the functions $G$ and ${\cal F}$ become
\begin{equation}
G_{1/2,3/4}(-z)=\frac{\sqrt{z}}\pi \exp \left( z^2/2\right) K_{1/4}\left(
\frac{z^2}2\right)  \label{Gs1}
\end{equation}
and
\begin{equation}
{\cal F}_{\nu ,1}\left( y\right) =\frac{2^{\nu +1}}{\sqrt{\pi }}\Gamma
\left( \nu +3/2\right) \frac y{\left( 1+y^2\right) ^{\nu +3/2}},
\label{Fs1}
\end{equation}
respectively. In order to obtain explicit results for the
amplitudes, numerical evaluations are unavoidable even in the
simplest case corresponding to zero temperature. In this case we
obtain from Eq.~(\ref{Xscaling})
\begin{eqnarray}\label{has1}
X_{{\rm ex}}^{{\rm u}} (0,0,0|1,1)&=&-\frac{\sqrt{y_0}}{8\pi^{\frac32}}
\sum_{\ell=1}^\infty\int_0^\infty dx
x^{-\frac32}\exp\left(\frac{y_0^2
x}{2}-\frac{\ell^2}{4x}\right)\nonumber\\ &&\times
K_{\frac14}\left(\frac{y_0^2 x}{2}\right)
-\frac{1}{3\pi}y_0^{3/2}.
\end{eqnarray}
Here $y_0$ is the solution of the equation for the spherical field
which can be obtained by requiring the partial derivative of the r.h.s
of (\ref{has1}) with respect to $y_0$ to be zero. Solving the last
equation numerically we end up with
$y_0=0.6248$.
After substitution of the solution in Eq.~(\ref{has1}) we obtain for
the Casimir amplitude
\begin{equation}
\Delta^u_{\rm Casimir}(0|1,1)=-0.3157.
\label{deltasigma1}
\end{equation}
The result given by Eq. (\ref{deltasigma1}) shows that the Casimir
amplitude in the case $\sigma=1$ is of larger magnitude than the
one in the case of short-range interaction. One can ask whether
this is just a coincidence or the Casimir amplitudes are
increasing function of $\sigma$ for a given fixed $d$.

\subsubsection{Relation with the Zamolochikov's $C$-function}

In terms of the critical-point value of an analog of the
finite-temperature $C$-function \cite{CN93}, the result given by Eq.
(\ref{eq14}) can be rewritten in the form
\begin{equation}
\Delta _{{\rm Casimir}}^{{\rm u}}\left( 0|d,2 \right) =
-n^u(d,2 )\tilde{c}^u(d,2),
\label{Sachdev}
\end{equation}
where in analogy with the  short-range interaction case above we
define a number ${\tilde c}^u $ via the relation
\begin{equation}
 {\tilde c}^u (d,\sigma)=
- \Delta^u_{\rm Casimir}(0|d,\sigma) / n^u(d,\sigma).
\label{cgen}
\end{equation}
Here, as usual, the normalization factor $n^u(d,\sigma)$ is chosen
so that to ensure ${\tilde c}^u = 1$ for the corresponding
Gaussian model, i.e.
\begin{equation}\label{nd}
n^u(d,\sigma)=\frac{2^{\sigma/2}\Gamma\left(\frac
d2+\frac\sigma4\right) \zeta\left(d+\frac\sigma2\right)}{\pi^{d/2}
\left | \Gamma\left(-\frac\sigma4 \right)\right |}.
\end{equation}
For $\sigma =2$ we immediately obtain $\tilde{c}^u(2,2)=4/5$,
where $n^u(d,2)=\Gamma ((d+1)/2)\zeta (d+1)/\pi ^{(d+1)/2}$
becomes the normalization factor given in \cite{CN93}. In that way
we reproduce the well known result for $\tilde{c}$ due to Sachdev
\cite{Sachdev93} who considered an example of a three dimensional
conformal field theory. This coincidence of the values of
$\tilde{c}$ is due to the fact that both models belong to the same
universality class. For more details on the behavior of the finite
temperature $C$-function in the case $\sigma=2,d=1,2,4$ see, e.g.
\cite{DT98}. The $d$-dependence of the value
$\tilde{c}^u(d,2)\equiv \Delta _{{\rm Casimir}}^{{\rm u}}\left(
0|d,2\right) /n^u(d,2)$ has been considered in \cite{PV981} for
$d$-dimensional ($2<d<4$) conformally invariant field theory. The
relation between the C function and the Casimir force for the
classical version of the model and short-range interaction has
been analyzed in some details for $d=2$ in~\cite{Danchev98}.

In the particular case of long-range interaction $d=\sigma=1$, in
accordance with Eq. (\ref{cgen}) and Eq.~(\ref{deltasigma1}) one gets
\begin{equation}
{\tilde c}^u(1,1)= 0.606.
\end{equation}

\subsection{``Temporal'' Casimir amplitudes}
\subsubsection{$\sigma$-dimensional system ($d=\sigma$)}
Let us note that unlike the case of "usual" Casimir amplitudes here it
is possible to consider the more general case $d=\sigma$, where
$0<\sigma\leq2$. From Eq. (\ref{xext}) one obtains a general
expression for the temporal Casimir amplitudes for a system with
geometry $\infty
^d\times L_\tau ,$ at the quantum critical point $\lambda=\lambda
_c$, $h=0,$
\begin{eqnarray}
&&\Delta _{{\rm Casimir}}^{{\rm t}}\left(d,\sigma \right) =
-\frac{k_d}{4\sqrt{\pi }\sigma }\Gamma \left(
\frac d\sigma \right) \Gamma
\left( -\frac {2d}\sigma -\frac12\right)y_0^{\frac dz+1}\nonumber \\
&&\ \ \ \ \ \ \ -\frac{k_d}{\sigma \sqrt{\pi }}\Gamma \left( \frac
d\sigma
\right)\left( 2y_0^2\right) ^{\frac d\sigma+
\frac12}
\sum_{m=1}^\infty\frac{K_{\frac d\sigma +\frac 12}\left( m y_0\right) }
{\left( m
y_0\right) ^{\frac d\sigma +\frac 12}},
\label{xexter}
\end{eqnarray}
where the scaling variable $y_0$ is the solution of the
corresponding equation for the spherical field. We notice here
that in the particular case $d/\sigma=1$ Eq.~(\ref{xexter})
simplifies considerably. That is why we are going to investigate
namely this case. In a way similar to that explained in the case
of short-range interactions, one obtains ($0<\sigma \leq 2$)
\begin{equation}
\Delta _{{\rm Casimir}}^{{\rm t}}(\sigma ,\sigma )=-\frac{16}{5\sigma }
\frac{\zeta (3)}{(4\pi )^{\sigma /2}}\frac 1{\Gamma (\sigma /2)}.
\label{eq16}
\end{equation}

Note that the defined ''temporal Casimir amplitude'' $ \Delta
_{{\rm Casimir}}^{{\rm t}}(\sigma ,\sigma )$ reduces for $\sigma
=2$ to the ''normal'' Casimir amplitude $\Delta _{{\rm
Casimir}}^{{\rm u}}\left( 0|2,2\right) $, given by
Eq.~(\ref{eq14}). This reflects the existence of a special
symmetry in that case between the ''temporal'' and the space
dimensionalities of the system.

When $\sigma \neq 2$ it is easy to verify that the following general
relation
\begin{equation}\label{dec}
\frac{\Delta _{{\rm Casimir}}^{{\rm t}}(\sigma ,\sigma )}{\Delta _{{\rm
Casimir}}^{{\rm t}}(2,2)}=\frac{8\pi}{\sigma(4\pi)^{\sigma/2}
\Gamma(\sigma/2)}
\end{equation}
between the temporal amplitudes holds. The r.h.s. of (\ref{dec})
is a decreasing function of $\sigma$.

\subsubsection{Relation with the Zamolochikov's $C$-function}
As it has been already mentioned in Section 2 if $z=1$ the
temporal excess free energy (see Eq. (\ref{deffext})) coincides,
up to a negative normalization constant, with the
nonzero-temperature generalization of the $C$-function of
Zamolodchikov proposed in \cite{CN93}. For $z\ne 1$ a
straightforward generalization of this definition can be proposed
at least in the case of long-range power-low decaying interaction
\begin{equation}
C(T,g|d,z)=-T^{-(1+d/z)} \frac {v^{d/z}}{n(d,z)} f_{{\rm ex}}^{{\rm t}}(T,g),
\label{Cgen}
\end{equation}
where $z=\sigma/2$, the nonuniversal constant $v$ in our notations is
$v=TL_{\tau}$ (see
Eq. (\ref{defxextgen})), and the normalization factor
is taken to be such
that the corresponding Gaussian model with the considered type of interaction
will have $C(T,g_c|d,z)\equiv {\tilde c}^t(d,\sigma )=1$ at its critical point, i.e.
\begin{equation}
n^t(d,\sigma )=\frac 4\sigma \frac{\zeta \left(1+2d/\sigma\right)}{(4\pi
)^{d/2}}\frac {\Gamma (2d/\sigma)} {\Gamma(d/2)}.
\label{nt}
\end{equation}

Let us note that the above choice of the normalization constant
$n^t(d,z)$ preserves not only the $\tilde c$ value for the
Gaussian model, but also the corresponding one for the spherical
model if $d=\sigma$. Indeed, in that case from (\ref{eq16}) and
the above definitions one again obtains that
\begin{equation}
\tilde c^t(\sigma,\sigma) =4/5
\label{ff}
\end{equation}
for the spherical model. The result given by Eq. (\ref{ff}) is a
generalization to the case of long-range interaction of the
Sachdev's result.

\section{Concluding remarks}
\label{Concl}

In the present article the free energy of a system with a geometry
$L\times \infty ^{d-1}\times L_\tau $ (where $\frac 12\sigma <d<\frac
32\sigma $), is derived (see Eq.~(\ref{eq8})). For $\sigma=2$ this
general result reduces to the one reported in~\cite{chamati97} where
only the case of short-range interactions has been considered. The
expression (\ref {eq10}) represents actually the verification of the
analog of the Privman-Fisher hypothesis~\cite{fisher84} for the
finite-size scaling form of the free energy (formulated initially for
classical systems) in the case when the quantum fluctuations are
essential. Note, that in that case one has a finite space dimension
and one additional finite dimension that is proportional to the
inverse temperature, which provide different types of critical regimes
and Casimir amplitudes. According to the finite-size scaling
hypothesis~\cite{fisher84,sachdev94} one has to expect that the
temperature multiplying the universal scaling function will be with
exponent $p=1+d/z$ \cite{sachdev94}, where the dynamic-critical
exponent $z$ reflects the anisotropic scaling between space and
``temporal'' (``imaginary-time'') directions.

Eqs.~(\ref{Xscaling}-\ref{eq12}) present a general expression for the
Casi\-mir force in the {\it quantum} spherical model. In the classical
limit ($\lambda
=0$) for a system with short-range interaction it coincides with the
corresponding one derived in~ \cite{Danchev96,Danchev98} for the
classical spherical model.

In order to derive the Casimir amplitudes in a simple analytical
closed form, some particular cases ($d=\sigma $) have been considered:

1) For the short-range case ($\sigma=2$) the corresponding amplitude
is given in Eq.~(\ref{eq14}). This amplitude is equal to the
''temporal Casimir amplitude'' for the ${\cal O}(n)$ sigma model in
the limit $n\to \infty $~\cite{Sachdev93}. In the short-range case we
have demonstrated {\it explicitly} that the two models, due to the
fact that they belong to the same universality class, indeed possess
equal Casimir amplitudes as it is to be expected on the basis of the
{\it hypothesis of universality}.

2) In the long-range case ($d=\sigma\ne 2$), the correction to the
ground-state energy of the bulk system due to the nonzero temperature
is determined by Eq.~(\ref{eq16}). One observes that in this case
$c^t(\sigma,\sigma)=4/5$ does not depend on $\sigma$. This could be
understood by noting that by changing $\sigma$ one does not change the
exponent in the spectrum that corresponds to the ``temporal''
(finite-size) dimensionality (see Eq.~(\ref{eq7})).

3) At zero temperature we evaluated numerically the Casimir amplitude
for the particular case $d=\sigma=1$ of the long-range interaction.
The result (\ref{deltasigma1}) shows that the Casimir amplitude is of
a larger magnitude than in the case of short range interaction
($\sigma=2$). Furthermore, the universal amplitude $c^u(1,1)$ is no
longer $\sigma$-independent, because the finite-size part of the
spectrum is $\sigma$-dependent in this case.

In accordance with the general expectations, all the amplitudes that
we have derived are {\it negative }(see Eqs. (\ref{Xcr}),
(\ref{eq14}), (\ref{eq16})) and (\ref{deltasigma1}).

Finally, let us note that the basic \ expression (see
Eqs.~(\ref{Xscaling})) for the scaling function of the excess free
energy can be used as a starting point for generalization of some of
the existing results on the $C$-function to the case of long-range
interactions. We have suggested in Eq. (\ref{Cgen}) a generalization
of the nonzero-temperature $C$-function, proposed by Neto and
Fradkin in \cite{CN93}, to the case of power-law long-range
interactions. For the quantum spherical model this definition leads
to $\tilde c =4/5$ for any $d=\sigma$ which generalizes to
long-range interactions the corresponding result for the case of
short-range interactions ($d=\sigma=2$) due to Sachdev
\cite{Sachdev93}.

\begin{acknowledgement}
This work is supported by The Bulgarian Science Foundation
(Projects F608/96 and MM603/96).

The authors would like to thank Jordan Brankov for stimulating
discussions.
\end{acknowledgement}

\appendix

\section{Mathematical Appendix}\label{appendixA}

First we explain how from Eq. (\ref{eq6}) one can obtain
(\ref{eq8}) for a system with a geometry $L\times \infty
^{d-1}\times L_\tau .$ It is easy to see that (\ref{eq6}) can be
rewritten in the form
\begin{eqnarray}
f(t,\lambda ,h;L)/J &=&\frac tN\sum_{{\bf q}}\ln \left\{ 2\sinh \left[ \frac
\lambda {2t}\sqrt{\phi +\frac{U({\bf q})}{J}}\right] \right\}\nonumber\\
& & -\frac{h^2}{2\phi }-\frac{\phi}{2} \nonumber\\
&=&-\frac{h^2}{2\phi }-\frac{\phi}{2}+A(L)-t\sum_{n=1}^\infty U_n(L),
\end{eqnarray}
where
\begin{equation}
A(L)=\frac \lambda {2N}\sum_{{\bf q}}\sqrt{\phi +|{\bf q}|^\sigma }
\end{equation}
and
\begin{equation}
U_n(L)=\frac 1N\sum_{{\bf q}}\exp \left[ -n\frac \lambda t\sqrt{\phi
+|{\bf q }|^\sigma }\right] .
\end{equation}
In order to calculate $A(L)$ we use the identity \cite{Chamati}
\begin{eqnarray}
\sqrt{1+z^\alpha }&=&\frac \alpha 2\int_0^\infty \left[ 1-\exp \left(
-zt\right) \right] t^{-1-\alpha /2}G_{\alpha ,1-\frac{\alpha}{2}}\left( -t^\alpha
\right)\nonumber\\
& &+1
\end{eqnarray}
(see Eq. (\ref{G}) for the definition of $G_{\alpha ,\beta }$).
Since we are interested in the geometry $L\times \infty
^{d-1}\times L_\tau $, one has by standard arguments
\begin{equation}
\frac 1N\sum_{{\bf q}}\rightarrow \frac 1{\left( 2\pi \right) ^{d-1}}\int
dq^{d-1}\times \frac 1L\sum_{q=-\left[ L-1\right] /2}^{\left[
L-1\right] /2}.
\label{vr}
\end{equation}
In the last sum over $q,$ by using the asymptotic formula
\cite{BT90}
\begin{eqnarray}
\frac 1L\sum_{q}\exp \left[
-a\left( \frac{2\pi q}L\right) ^2\right] &&\approx \frac 1{\sqrt{4\pi a}
}\left\{ {\rm erf}\left( \pi \sqrt{a}\right) \right.\nonumber\\
&&\left.+2\sum_{l=1}^\infty \exp
\left[ -\frac{l^2L^2}{4a}\right] \right\} ,
\end{eqnarray}
valid for $L>>1$, we obtain, after some algebra,
\begin{equation}
A(L)=A(\infty )+\delta A(L),
\end{equation}
where
\begin{equation}
A(\infty )=\frac \lambda 2\int_{-\pi }^\pi dq_1\ldots \int_{-\pi }^\pi
dq_d\sqrt{\phi +\left( q_1^2+\ldots +q_d^2\right) ^{\sigma /2}}
\end{equation}
and \cite{Chamati}
\begin{eqnarray}
\delta A(L)&=&-\frac \lambda 4\frac \sigma {\left( 4\pi \right)
^{d/2}}\sum_{l=1}^\infty \int_0^\infty x^{-\sigma /4-d/2-1}\exp \left( -
\frac{l^2L^2}{4x}\right)\nonumber\\
& & \times G_{\frac \sigma 2,1-\frac \sigma 4}\left(
-x^{\sigma /2}\phi \right) dx.
\end{eqnarray}
Since the only singularities of $A(\infty )$ as a function of
$\phi $ are coming from small $q$'s, it is justified to use a
sphericalization of the Brillouin zone, which leads to
\begin{eqnarray}
A(\infty ) &\simeq &\frac \lambda
2k_d\int_0^{x_D}\frac{dx}{x^{1-d}}\sqrt{\phi +x^\sigma } \\ &\simeq
&\frac{\lambda k_d}{2d}x_D^d\left( x_D^\sigma +\phi \right)
^{1/2}\nonumber\\
& & \times _2F_1\left( 1,-\frac 12,1+\frac d\sigma ,\frac{x_D^\sigma
}{ x_D^\sigma +\phi }\right) ,
\end{eqnarray}
where $_2F_1$ is the hypergeometric function. Now it is clear how
the ``first half'' of (\ref{eq8}) can be obtained. Next we turn to
evaluation of the term $U_n(L).$ Taking into account (\ref{vr}),
we rewrite $U_n(L)$ in the form
\begin{eqnarray}
U_n(L)&=&\frac{L^{-1}}{\left( 2\pi \right) ^{d-1}}
\sum_{q=-(L-1)/2}^{(L-1)/2}\int_{-\pi }^\pi dq_2\ldots \int_{-\pi
}^\pi dq_d\nonumber\\ & &\times\exp \left[ -n\frac \lambda
t\sqrt{\phi +\left( q_1^2+\ldots +q_d^2\right)
^{\frac\sigma2}}\right].
\end{eqnarray}
Using the Poisson summation formula
\begin{eqnarray}
\sum_{n=a}^bf\left( n\right) &=&\sum_{k=-\infty }^\infty  \int_a^bdn\exp \left[
i2\pi kn\right] f\left( n\right)\nonumber\\
& & +\frac 12\left[ f\left( a\right) +f\left(
b\right) \right]
\end{eqnarray}
we obtain from the above expression
\begin{equation}
U_n(L)=U_n(\infty )+\delta U_n(L),
\end{equation}
where
\begin{equation}
U_n(\infty )=k_d\int_0^{x_D}dxx^{d-1}\exp \left[ -n\frac \lambda t\sqrt{\phi
+x^\sigma }\right]  \label{ui}
\end{equation}
and
\begin{eqnarray}
\delta U_n(L)&=&\frac 2{\left( 2\pi \right) ^d}\sum_{l=1}^\infty \int_{-\pi
}^\pi dq_1\ldots \int_{-\pi }^\pi dq_d \cos \left[
q_1lL\right]\nonumber\\ & &\times\exp \left[ -n\frac \lambda t\sqrt{
\phi +\left( q_1^2+\ldots +q_d^2\right) ^{\sigma /2}}\right]  .  \label{ul}
\end{eqnarray}
In the low-temperature limit $t<<1$ one can replace in (\ref{ui})
$x_D$ by infinity. Then, using the integral representation of
$K_\nu (x)$
\begin{eqnarray}
&&K_\nu (2\sqrt{zt})=K_{-\nu }(2\sqrt{zt})
\nonumber\\
&&\ \ \ \ =\frac 12\int_0^\infty \left( \frac zt\right)
^{\frac{\nu}{2}}x^{-\nu
-1}\exp
\left(-t x-\frac zx\right) dx
\end{eqnarray}
we derive
\begin{eqnarray}
U_n(\infty )&=&\frac \lambda {t\sigma }\frac{k_d}{\sqrt{\pi }}\Gamma \left(
\frac d\sigma \right) \phi ^{\frac d\sigma +\frac 12}K_{\frac d\sigma +\frac
12}\left( n\frac \lambda t\phi ^{\frac 12}\right) \nonumber\\
&&\times\left( n\frac \lambda
{2t}\phi ^{\frac 12}\right) ^{-\left( \frac d\sigma +\frac 12\right) }.
\label{uif}
\end{eqnarray}
We are left to deal now only with $\delta U_n(L).$ Sphericalizing
the Brillouin zone in (\ref{ul}), performing the integrations, by
using the integral representation for the Bessel function $J_\nu
(z)$
\begin{eqnarray}
\int_{-a}^a&&\left( a^2-x^2\right) ^{\beta -1}\exp \left[ i\lambda x\right] dx
\nonumber\\
& & \ \ \ \ \ \ \
=\sqrt{\pi }\Gamma \left( \beta \right) \left( \frac{2a}\lambda \right)
^{\beta -1/2}J_{\beta -1/2}(a\lambda ),
\end{eqnarray}
( ${\rm Re} \ \beta >0$) we get
\begin{eqnarray}
\delta U_n(L)&=&\frac{2L^{-d/2+1}}{\left( 2\pi \right) ^{d/2}}
\sum_{l=1}^\infty \int_0^{x_D}\frac{x^{d/2}}{l^{d/2-1}}J_{d/2-1}\left(
lLx\right) \nonumber\\
&&\times\exp \left[ -n\frac \lambda t\sqrt{\phi +x^\sigma }\right] .
\label{ulf}
\end{eqnarray}
In the low-temperature limit the upper limit of integration in the
above expressions can be replaced by infinity. From (\ref{uif})
and (\ref{ulf}) one obtains the last two terms in Eq. (\ref{eq8}).


\begin{thebibliography}{99}
\bibitem{note1} In that case  both finite-size and van der Waals
forces give raise to a contribution to the free energy of the
wetting layer that depends on its thickness $L$ as $L^{-2}$ for
$d=3$.
\bibitem{note2} The requirement the finite nonzero-temperature system
to be in ordered state involves $d-1>\sigma$, which together
 with $\frac 12 \sigma<d<\frac 32 \sigma$ excludes the case $\sigma=2$.
\end{thebibliography}
\end{document}